\documentclass[a4paper,fleqn]{cas-sc}

\usepackage[numbers]{natbib}
\usepackage{amssymb,amsmath}
\usepackage{slashed}
\usepackage{bm}

\def\tsc#1{\csdef{#1}{\textsc{\lowercase{#1}}\xspace}}
\tsc{WGM}
\tsc{QE}

\DeclareMathOperator{\nrdi}{\overleftrightarrow{d}}
\DeclareMathOperator{\nrp}{\overleftrightarrow{\partial}}

\begin{document}
\let\WriteBookmarks\relax
\def\floatpagepagefraction{1}
\def\textpagefraction{.001}

\shorttitle{${\bf P}$-dependent NN interaction}    

\shortauthors{E. Filandri and L. Girlanda}  

\title [mode = title]{Momentum dependent nucleon-nucleon contact interaction from a relativistic Lagrangian}

\author[1,2]{E. Filandri}[
       orcid=0000-0001-7263-1491]


\affiliation[1]{organization={Department of Mathematics and Physics, University of Salento},
            addressline={Via per Arnesano}, 
            city={Lecce},
            postcode={I-73100}, 
            country={Italy}}

\affiliation[2]{organization={INFN, Sez. di Lecce},
            city={Lecce},
            postcode={I-73100}, 
            country={Italy}}

\author[1,2]{L. Girlanda}[
       orcid=0000-0002-5560-005X]


\cormark[1]

\cortext[1]{Corresponding author}

\begin{abstract}
  A complete set of parity- and time-reversal conserving relativistic nucleon-nucleon contact operators is identified up to the order $O(p^4)$ of the expansion in soft momenta $p$. A basis is also provided for the corresponding non-relativistic operators contributing in the general reference frame. We show that the non-relativistic expansions of the relativistic operators involve twenty-six independent combinations, two starting at $O(p^0)$, seven at order $O(p^2)$ and seventeen at order $O(p^4)$. This gives supporting evidence to the existence of two free low-energy constants which parametrize an interaction depending on the total nucleon pair momentum ${\bf P}$, and were recently found to be instrumental for the resolution of the long standing $A_y$ problem in low-energy $p-d$ scattering. Furthermore, all remaining ${\bf P}$-dependent interactions at the same order are uniquely determined as relativistic corrections.

\end{abstract}


\maketitle

\section{Introduction}

The systematic approach provided by Effective Field Theories (EFTs) has become the standard framework to address the nuclear interaction problem \cite{Weinberg:1990rz,Weinberg:1991um,Weinberg:1992yk,Bedaque:2002mn,Epelbaum:2005pn,Epelbaum:2008ga,Machleidt:2011zz}. Its starting point is the identification of the most general effective Lagrangian respecting all the relevant symmetries, e.g. the approximate chiral symmetry of QCD, and on a power counting to order the infinite tower of allowed interactions. As a result, a predictive setting emerges in which physical observables are expressed at each order of the low-energy expansion in terms of a finite number of low-energy constants (LECs) which can be treated as fitting parameters and determined from phenomenology.
The precise identification of the necessary and sufficient number of  such parameters at each order of the expansion is an important task, not only to put the theory to a stringent test, but also for the sake of a robust estimate of the theoretical uncertainty from neglected orders \cite{Furnstahl:2014xsa,Epelbaum_2015,Furnstahl:2015rha,Konig:2019adq}.
In the context of the nuclear forces, these fitting parameters are the LECs associated with contact interactions among nucleons, which are left unconstrained by chiral symmetry. However, they are still constrained by the Poincar\'e symmetry \cite{Foldy:1960nb,Krajcik:1974nv}. The latter has to be recovered eventually, despite the fact that the usual setting for nuclear physics is a non-relativistic quantum-mechanical one. Relativistic effects coming from different sources, like recoil corrections in energy denominators and vertex corrections from the heavy baryon expansion \cite{Jenkins:1990jv} matched to the relativistic theory \cite{Bernard:1992qa}, or equivalently with reparametrization invariance \cite{Luke:1992cs}, must in the end conspire to satisfy such constraints.
Since relativistic effects scale with the soft nucleon momenta, they can be examined order by order in the low-energy expansion and the constraints on the interactions imposed algebraically \cite{Girlanda:2010ya}.
The analysis of such constraints in Ref.~\cite{Girlanda:2020pqn} has lead to the identification of two free LECs, named $D_{16}$ and $D_{17}$  parametrizing a nucleon-nucleon (NN) interaction depending on the overall pair momentum  ${\bf P}$ at the same order as the ${\bf P}$-independent interaction parametrized by the 15 LECs $D_{1,...,15}$ introduced in Ref.~\cite{Entem:2001cg,Entem:2003ft}.
As a matter of fact, this remarkable outcome was not so much about the restrictions on existing LECs, but rather about the existence of unconstrained LECs.
Furthermore, in Ref.~\cite{Girlanda:2023znc} the two extra LECs where found to be crucial for the resolution of the long-standing $A_y$ puzzle of low-energy $p-d$ scattering, which persisted even with the more advanced models of three-nucleon interactions \cite{Golak:2014ksa}.

The purpose of the present paper is to exhibit supporting evidence to this important result starting from a manifestly Lorentz invariant NN contact Lagrangian density and performing the non-relativistic reduction up to order $1/m^4$, $m$ being the nucleon mass.
Indeed, besides being interesting for its own sake (see e.g. Ref.~\cite{Epelbaum:2012ua}), a relativistically invariant effective Lagrangian guarantees that all requirements of relativity be automatically satisfied as one proceeds systematically in the low-energy expansion. 
A similar path was followed in Ref.~\cite{Xiao:2018jot}. However, in that paper the analysis was conducted in the NN center-of-mass system, preventing to analyze the ${\bf P}$-dependent operators we are interested in.
The plan of the paper is the following. In Section~\ref{sec:2} an overcomplete set of relativistic invariant NN contact operators is obtained up to order $p^4$ of the low-energy expansion.
In Section~\ref{sec:3} a basis of non-relativistic contact operators at the same order  is identified as consisting of 45 members respecting isospin, parity and time reversal symmetry,  15 of which are ${\bf P}$-independent.
In Section~\ref{sec:4}  we derive the non-relativistic expansion of the Lorentz invariant operators in terms of the adopted basis of non-relativistic operators and identify the linearly independent combinations as consisting of 2 operators at $O(p^0)$, 7 at $O(p^2)$ and 17 at $O(p^4)$, thus confirming the results of Ref.~\cite{Girlanda:2020pqn}. We also determine the remaining 28 $O(p^4)$ LECs as $1/m$ corrections to the lower-order interactions.  Finally, we draw the conclusions in Section~\ref{sec:5} .

\section{Relativistic contact operators}\label{sec:2}
The general expression of the relativistic  NN contact Lagrangian  may be derived along the lines of Ref.~\cite{Girlanda:2010zz, Petschauer:2013uua}. It consists of products of fermion bilinears such as
\begin{equation}
  (\bar{\psi}\overleftrightarrow{\partial}_{\mu_1}\cdots\overleftrightarrow{\partial}  _{\mu_i}\Gamma_A\psi)\partial_{\lambda_1}\cdots\partial_{\lambda_k}(\bar{\psi}\overleftrightarrow{\partial}_{\nu_1}\cdots\overleftrightarrow{\partial}  _{\nu_j}\Gamma_B\psi)\,,
\end{equation}
where $\psi$ indicates the relativistic nucleon field, a doublet in isospin space, $\overleftrightarrow{\partial}=\overrightarrow{\partial}-\overleftarrow{\partial}$ and $\Gamma_{A,B}$ are generic elements of the Clifford algebra including the metric and Levi-Civita tensor $\epsilon^{\mu\nu\rho\sigma}$ with the convention $\epsilon^{0123}=-1$. The Lorentz indices on the partial derivatives must be contracted  among themselves and/or with those in the $\Gamma_{A,B}$ in order to fulfill the Lorentz invariance.
To construct the covariant Lagrangian, also isospin, parity, charge conjugation and time reversal symmetry must be satisfied, in addition to the requirement of hermiticity \cite{Fettes:1998ud,Fettes:2000gb}.
Local chiral symmetry is not our concern here, but it could be enforced by promoting the derivatives to chiral covariant ones.
In Table \ref{tab:Trans.Pr} we list the transformations proprieties of the fermion bilinears under the discrete symmetries.
  According to the CPT theorem, time reversal symmetry is automatically satisfied if charge conjugation and parity symmetries are fulfilled.

\begin{table}
\caption{Transformation proprieties of the differents elements of the Clifford algebra, metric tensor, Levi-Civita tensor and derivative operators under parity ($\mathcal{P}$), charge conjugation ($\mathcal{C}$) and hermitian coniugation (h.c.)}\label{tab:Trans.Pr}
\begin{center}
\begin{tabular*}{\tblwidth}{@{}LLLLLLLLLL@{}}
\toprule
&$1$&$\gamma_5$&$\gamma_\mu$&$\gamma_\mu\gamma_5$&$\sigma_{\mu\nu}$&$g_{\mu\nu}$&$\epsilon_{\mu\nu\rho\sigma}$&$\overleftrightarrow{\partial}_\mu$&$\partial_\mu$\\
\midrule
$\mathcal{P}$&$+$&$-$&$+$&$-$&$+$&$+$&$-$&$+$&$+$\\
$\mathcal{C}$&$+$&$+$&$-$&$+$&$-$&$+$&$+$&$-$&$+$\\
h.c.&$+$&$-$&$+$&$+$&$+$&$+$&$+$&$-$&$+$\\
\bottomrule
\end{tabular*}
\end{center}
\end{table}
Regarding the isospin degrees of freedom,   both flavour structures  $1\otimes1$ and $\tau^a\otimes \tau^a$ are allowed. However the latter can  be disregarded since it can be eliminated by Fierz rearrangements in the case of identical particles. Alternatively, since in the two-nucleon system the isospin quantum number is fixed in terms of the orbital and spin quantum numbers, we can equally well disregard it and treat the nucleons as distinguishable.
To specify the chiral order of each building block, it is necessary to identify the powers of soft nucleon momenta $p$.
The derivatives $\partial$ acting on the entire bilinear is of order $p$, while the derivative $\overleftrightarrow{\partial}$ acting inside a bilinear is of $p^0$  due to the presence of the heavy fermion  mass scale. Therefore one could have at each order any number of $\nrp$. However, whenever a four-gradient Lorentz index is contracted with one of the elements of the Clifford algebra, the fields' equations of motion can be used to remove it \cite{Georgi:1991ch,Arzt:1993gz}. Furthermore, by the same equations of motion,
$\nrp^\mu \nrp_\mu = 4 m^2 - \partial^2$. 
As a result, no two Lorentz indices inside a fermion bilinear can be contracted with one another, except for the Levi-Civita
tensors and for the suppressed $\partial^2$.
But we still have in principle any numbers $n$ of pairwise contracted $(\nrp_A \cdot \nrp_B)^n$ between the two different bilinears. However, as observed in Ref.~\cite{Xiao:2018jot}, we can limit  ourselves to terms with $n= 0, 1, 2$  up to $p^4$. Indeed, considering the initial (final) four-momenta $p_1, p_2$ ($p'_1, p'_2$), instead of space-time derivatives, and using the mass shell
relation for the momenta one has\footnote{Such equalities are always understood as valid up to terms which vanish by the equations of motion.}
\begin{equation}
\nrp_A\cdot \nrp_B = -4 m^2 +\frac{1}{2} \left[ (p_1-p_2)^2 + (p_1 -p_2')^2 + (p_1'-p_2)^2 + (p_1'-p_2')^2 \right] = -4 m^2 + O(p^2).
\end{equation}

Further criteria in specifying the power counting of the operators regard the Dirac matrix $\gamma_5$, which can be thought of as $O(p)$ since it mixes the large and small components of the Dirac spinor, and the Levi-Civita tensor $\epsilon_{\mu\nu\rho\sigma}$, which, when contracted with $n$ derivatives acting inside a bilinear, raises the chiral order by $n-1$. 
These criteria lead to the complete (although non-minimal) set of 
 relativistic contact operators displayed in Table \ref{tab:RelOp}. The last column contains, for each one of the Dirac structures, the additional combination of four-gradients arising up to $O(p^4)$.
Each operator to which the subscripts refer is to be understood multiplied by its own factor, e.g.  $4 m^2 O_3=\bar{\psi} \psi \partial^2 \bar{\psi} \psi$. 
Among the operators of the class $1\otimes\gamma$ it is sufficient to consider only the first three combinations of four-gradients, since they are equal to $O_1$ at $O(p^0)$ and then, e.g.
\begin{equation}
    \frac{i}{2 m} \bar \psi \nrp^\mu \psi \partial^4 \bar \psi \gamma_\mu \psi = \bar \psi \psi \partial^4 \bar \psi \psi + O(p^6).
\end{equation}
 The same reasoning holds  for $O_{17-19}$. Analogously for the operators $O_{43-45},\,O_{46-48}$ and $O_{49-51}$ since the quartic terms are contained in $O_{34-36}$.

Notice that $\partial$, the four-gradient acting on the whole bilinear, is always thought of as acting on the second bilinear. Indeed one can use (four-dimensional) partial integrations and always reduce to this case. Such partial integrations leave of course the action invariant. Therefore, eventhough affecting  the Hamiltonian density  off the energy shell, this procedure does not change the physical observables.

Our construction differs from the one conducted in Ref.~\cite{Xiao:2018jot}, due to a different choice of operators reduced by the equations of motion, e.g.  terms with $\epsilon^{\mu\nu\alpha\beta}$ are  eliminated. 

\begin{table}
\caption{A complete non-minimal set of relativistic contact operators. Here $\nrdi_X=\nrp_X/(2 m)$ ($X=A,B$) and $d=\partial/(2 m)$ denote the adimensional four-gradients.}\label{tab:RelOp}
  \begin{center}
\begin{tabular*}{\tblwidth}{@{}LLL@{}}
\toprule
$\Gamma_A\otimes\Gamma_B$ & Operators & Gradient structures \\
    \midrule
$1 \otimes  1$& $\tilde{O}_{1-6} = \bar{\psi} \psi\bar{\psi} \psi$ & $\{1, \nrdi_A\cdot \nrdi_B, d^2, (\nrdi_A\cdot \nrdi_B)^2 , (\nrdi_A\cdot \nrdi_B) d^2, d^4\}$\\ 
\hline 
$1\otimes \gamma$ &  $\tilde{O}_{7-9} = \frac{i}{2m}\bar{\psi}\overleftrightarrow{\partial}^\mu \psi \bar{\psi} \gamma_\mu \psi$&  $\{1,\nrdi_A\cdot\nrdi_B, d^2\}$ \\ 
\hline 
$1 \otimes\gamma\gamma_5$ &$\tilde{O}_{10-12} = \frac{-1}{8m^3} \epsilon^{\mu\nu\alpha\beta} \bar{\psi}\overleftrightarrow{\partial}_\mu\psi\partial_\nu\bar{\psi}\gamma_\alpha\gamma_5\overleftrightarrow{\partial}_\beta\psi$ & $\{1,\nrdi_A\cdot\nrdi_B, d^2\}$\\ 
\hline 
$\gamma_5\otimes\gamma_5$& $\tilde{O}_{13-15} = -\bar{\psi}\gamma_5 \psi\bar{\psi}\gamma_5 \psi$ & $\{1,\nrdi_A\cdot\nrdi_B, d^2\}$\\ 
\hline 
$\gamma_5\otimes\sigma  $ & $\tilde{O}_{16} = \frac{1}{16m^4}\epsilon^{\mu\nu\alpha\beta} \bar{\psi}\gamma_5\overleftrightarrow{\partial}^\gamma\overleftrightarrow{\partial}_\mu\psi\partial_\nu\bar{\psi}\sigma_{\alpha\gamma}\overleftrightarrow{\partial}_\beta\psi$&$1$\\ 
\hline 
$\gamma\otimes\gamma$ & $\tilde{O}_{17-19} = \bar{\psi} \gamma_\mu\psi\bar{\psi}\gamma^\mu \psi$  &  $\{1,\nrdi_A\cdot\nrdi_B, d^2\}$ \\ 
 &$\tilde{O}_{20-22} = \frac{-1}{4m^2}\bar{\psi} \gamma_{\mu}\overleftrightarrow{\partial}_{\nu}\psi\bar{\psi}\gamma^{\nu}\overleftrightarrow{\partial}^{\mu} \psi$  &  $\{1,\nrdi_A\cdot\nrdi_B, d^2\}$ \\ 
\hline 
$\gamma\otimes\gamma\gamma_5$  & $\tilde{O}_{23} = \frac{-i}{16m^4} \epsilon^{\mu\nu\alpha\beta}\bar{\psi}\gamma_\mu\overleftrightarrow{\partial}^\gamma\overleftrightarrow{\partial}_\nu\psi\partial_\alpha\bar{\psi}\gamma_\gamma\gamma_5\overleftrightarrow{\partial}_\beta\psi$&  $1$ \\ 
 &   $\tilde{O}_{24} = \frac{-i}{16m^4} \epsilon^{\mu\nu\alpha\beta}\bar{\psi}\gamma^\gamma\overleftrightarrow{\partial}_\mu \psi\partial_\nu\bar{\psi}\gamma_\alpha\gamma_5\overleftrightarrow{\partial}_\gamma\overleftrightarrow{\partial}_\beta \psi$&$1$ \\  
 & $\tilde{O}_{25-27} = \frac{i}{4m^2} \epsilon^{\mu\nu\alpha\beta}\bar{\psi}\gamma_\mu\overleftrightarrow{\partial}_\nu \psi\partial_\alpha\bar{\psi}\gamma_\beta\gamma_5 \psi$ & $\{1,\nrdi_A\cdot\nrdi_B, d^2\}$  \\ 
 & $\tilde{O}_{28-30} = \frac{i}{4m^2} \epsilon^{\mu\nu\alpha\beta}\bar{\psi}\gamma_\mu\psi\partial_\nu\bar{\psi}\gamma_\alpha\gamma_5\overleftrightarrow{\partial}_\beta \psi$ &$ \{1,\nrdi_A\cdot\nrdi_B, d^2\}$  \\ 
\hline 
$\gamma\gamma_5\otimes\gamma\gamma_5$ & $\tilde{O}_{31-36} = \bar{\psi}\gamma^\mu\gamma_5\psi\bar{\psi}\gamma_\mu\gamma_5\psi$ & $\{1, \nrdi_A\cdot \nrdi_B, d^2, (\nrdi_A\cdot \nrdi_B)^2, (\nrdi_A\cdot \nrdi_B) d^2, d^4\}$\\ 
& $\tilde{O}_{37-39} = \frac{-1}{4m^2}\bar{\psi}\gamma^\mu\gamma_5\overleftrightarrow{\partial}^\nu\psi\bar{\psi}\gamma_\nu\gamma_5\overleftrightarrow{\partial}_\mu\psi$ & $\{1,\nrdi_A\cdot\nrdi_B, d^2\}$ \\ 
\hline 
$\gamma\gamma_5\otimes\sigma$ &  $\tilde{O}_{40-42} = \frac{-i}{8m^3}\epsilon^{\mu\nu\alpha\beta}\bar{\psi}\gamma_\mu\gamma_5\overleftrightarrow{\partial}_\nu\overleftrightarrow{\partial}^\gamma\psi\bar{\psi}\sigma_{\alpha\gamma}\overleftrightarrow{\partial}_\beta\psi$ & $\{1,\nrdi_A\cdot\nrdi_B, d^2\}$  \\ 
 & $\tilde{O}_{43-45} = \frac{i}{2m}\epsilon^{\mu\nu\alpha\beta}\bar{\psi}\gamma_\mu\gamma_5\psi\bar{\psi}\sigma_{\nu\alpha}\overleftrightarrow{\partial}_\beta\psi$ &  $\{1,\nrdi_A\cdot\nrdi_B, d^2\}$ \\ 
  & $\tilde{O}_{46-48} = \frac{i}{2m}\epsilon^{\mu\nu\alpha\beta}\bar{\psi}\gamma_\mu\gamma_5\overleftrightarrow{\partial}_\nu \psi\bar{\psi}\sigma_{\alpha\beta}\psi$ &  $\{ 1,\nrdi_A\cdot\nrdi_B, d^2 \}$ \\ 
\hline 
$\sigma\otimes\sigma$ & $\tilde{O}_{49-51} = \bar{\psi}\sigma_{\mu\nu}\psi\bar{\psi}\sigma^{\mu\nu}\psi$ & $\{1,\nrdi_A\cdot\nrdi_B, d^2\}$ \\ 
 & $\tilde{O}_{52-54} = \frac{-1}{4m^2}\bar{\psi}\sigma^{\mu\alpha}\overleftrightarrow{\partial}^\beta\psi\bar{\psi}\sigma_{\mu\beta}\overleftrightarrow{\partial}_\alpha\psi$ & $\{1,\nrdi_A\cdot\nrdi_B, d^2\}$  \\ 
\hline
\end{tabular*} 
\end{center}
\end{table}
\section{Non-relativistic operator basis in the general reference frame}\label{sec:3}
The purpose of this section is to establish a minimal basis of non-relativistic NN contact operators involving 4 powers of three-gradients, i.e. at order $O(p^4)$, in terms of non relativistic nucleon fields $N(x)$, an isospin doublet.
It is well known that, in the NN center-of-mass system, such a basis consists of 15 operators \cite{Entem:2001cg,Entem:2003ft}, while in the general frame, to the best of our knowledge, it was never considered.
The steps for constructing it are similar to the relativistic case:
parity invariance requires an  even number of gradient operators, and
time reversal symmetry and hermiticity must also be respected.
As in the relativistic case, regarding the isospin structure, we can limit ourselves to the $1\otimes 1$ structure, since the corresponding $\tau^a\otimes\tau^a$ structures can be obtained through Fierz rearrangements.
Using the above criteria, we arrive at a list of 45 operators in the general frame, displayed in Table \ref{tab:NRBase} through the corresponding  matrix elements between initial and final nucleons having momenta ${\bf p}_1, {\bf p}_2$ and  ${\bf p'}_1, {\bf p'}_2$ respectively, where
$\bm{ k}=\bm{ p}'-\bm{ p}$, $\bm{ Q}=\frac{\bm{ p}'+\bm{ p}}{2}$, with $\bm{ p}=({\bf p}_1- {\bf p}_2)/2$   and $\bm{ p'}=({\bf p'}_1- {\bf p'}_2)/2$ the initial and final relative momenta respectively. In the same table, for completeness, the $O(p^0)$ and $O(p^2)$ bases of operators are also shown.
The choice of the ${\bf P}$-independent operators conforms to the standard convention for the corresponding LECs \cite{Entem:2003ft} defining the NN contact potential in the center-of-mass frame. 
Thus the NN contact Hamiltonian density takes the form\begin{equation}
    {\cal H}_{NN} = {\cal H}^{(0)} + {\cal H}^{(2)} + {\cal H}^{(4)},
    \end{equation}
    with
    \begin{align}
        {\cal H}^{(0)}&=C_S O_S +  C_T O_T, \label{eq:h0}\\ 
            {\cal H}^{(2)} &=  \sum_{i=1}^7 C_i O'_i +  \sum_{i=1}^5 C_i^* O^{*\prime}_i, \label{eq:h2}\\
            {\cal H}^{(4)} &=  \sum_{i=1}^{17} D_i O''_i +  \sum_{i=1}^{28} D_i^* O^{*\prime\prime}_i \label{eq:h4}.
\end{align}
The additional LECs, $D_{16}$, $D_{17}$, $C^*_i$ and $D^*_i$ parametrize the ${\bf P}$-dependent NN interacton in the general reference frame and do not contribute in the center-of-mass frame. As will be clear in the following, the former two, already identified in Ref.~\cite{Girlanda:2020pqn}, are unconstrained by relativity, while the $C_i^*$ and $D^*_i$ are fixed and represent drift corrections to the lower-order interactions \cite{Forest:1995sg}.
Notice also that 
operators like ${\bf k}\times {\bf Q}\cdot {\bf P}\, {\bf Q}\cdot({\bm \sigma}_1-{\bm \sigma}_2)$ can be expressed in the chosen basis through the identity
\begin{equation}
    ({\bf A}\times {\bf B}\cdot {\bf C}) {\bf D}\cdot{\bm \sigma} ={\bf A}\cdot {\bf D} ({\bf B}\times {\bf C}\cdot{\bm \sigma})- {\bf B} \cdot {\bf D} ({\bf A}\times {\bf C}\cdot {\bm \sigma} )+ {\bf C}\cdot {\bf D} ( {\bf A} \times {\bf B}\cdot{\bm \sigma}).
\end{equation}

\begin{table*}[width=.9\textwidth,cols=4,pos=h]
\caption{A complete basis of non-relativistic operators computed between states of two nucleons with initial (final) momenta ${\bf p}_1 ({\bf p}'_1) ={\bf P}/2 + {\bf Q} \pm {\bf k}/2$ and ${\bf p}_2 ({\bf p}'_2)= {\bf P/2} - {\bf Q} \pm {\bf k}/2$.  Here $\gamma_{S/T}= C_{S/T}/(4 m^2)$.}\label{tab:NRBase}
\begin{scriptsize}
\begin{tabular*}{\tblwidth}{@{} LLLL@{} }
   \toprule
Operator & LEC & Operator & LEC $\times (4 m^2)$ \\
   \midrule
$O_S = 1$ & $C_S$ & $O^{*\prime\prime}_{1} = ik^2(\bm{k}\times\bm{P}\cdot(\bm{\sigma}_1-\bm{\sigma}_2))$ &
$D_1^*=\frac{3}{8} (\gamma_S - \gamma_T)   - \frac{1}{2}( C_1 - C_3)$ \\
$O_T = \bm{\sigma}_1\cdot\bm{\sigma}_2$ & $C_T$ & $O^{*\prime\prime}_{2} = \bm{k}\cdot\bm{P}(\bm{k}\cdot\bm{\sigma}_1\,\bm{Q}\cdot\bm{\sigma}_2-\bm{k}\cdot\bm{\sigma}_2\,\bm{Q}\cdot\bm{\sigma}_1)$
& $D_2^*=\frac{1}{4} C_5  +  C_6  $ \\
\cline{1-2}
$O'_{1} = k^2$ & $C_1$ & $O^{*\prime\prime}_{3} = k^2(\bm{Q}\cdot\bm{\sigma}_1\,\bm{P}\cdot\bm{\sigma}_2-\bm{Q}\cdot\bm{\sigma}_2\,\bm{P}\cdot\bm{\sigma}_1)$
& $ D_3^*=\frac{3}{2} \gamma_T - 2 C_3 +  \frac{1}{4} C_5 $ \\
$O'_{2} =  Q^2$ & $C_2$ & $O^{*\prime\prime}_{4} = k^2P^2$
& $D_4^*=\frac{1}{4} (3 \gamma_S + \gamma_T)  -C_1 $ \\
$O'_{3} =  k^2 \bm{\sigma}_1\cdot\bm{\sigma}_2 $ & $C_3$ & $O^{*\prime\prime}_{5} = k^2P^2\bm{\sigma}_1\cdot\bm{\sigma}_2$
& $D_5^*=\frac{1}{4} (\gamma_S + 3 \gamma_T) -C_3 $ \\
$O'_{4} = Q^2 \bm{\sigma}_1\cdot\bm{\sigma}_2$ & $C_4$ & $O^{*\prime\prime}_{6} = (\bm{k}\cdot\bm{P})^2$
& $ D_6^*=\frac{1}{4} (5 \gamma_S - \gamma_T) -C_1  $ \\
$O'_{5} = i\frac{\bm{\sigma}_1+\bm{\sigma}_2}{2}\cdot\bm{Q}\times\bm{k}$ & $C_5$ & $O^{*\prime\prime}_{7} = (\bm{k}\cdot\bm{P})^2\bm{\sigma}_1\cdot\bm{\sigma}_2$
& $ D_7^*=-\frac{1}{4} (\gamma_S - 5 \gamma_T)  -C_3 $ \\
$O'_{6} = \bm{\sigma}_1\cdot\bm{k}\bm{\sigma}_2\cdot\bm{k}$ & $C_6$ & $O^{*\prime\prime}_{8} = P^2\,\bm{k}\cdot\bm{\sigma}_1\,\bm{k}\cdot\bm{\sigma}_2$
& $ D_8^*=-\frac{1}{4} (\gamma_S -\gamma_T)  -C_6 $ \\
$O'_{7} = \bm{\sigma}_1\cdot\bm{Q}\bm{\sigma}_2\cdot\bm{Q} $& $C_7$ & $O^{*\prime\prime}_{9} = \bm{k}\cdot\bm{P}(\bm{k}\cdot\bm{\sigma}_1\,\bm{P}\cdot\bm{\sigma}_2+\bm{k}\cdot\bm{\sigma}_2\,\bm{P}\cdot\bm{\sigma}_1)$
& $D_9^*=\frac{1}{4} (\gamma_S - \gamma_T) -\frac{1}{2} C_6  $ \\
$O^{*\prime}_{1} = i\frac{\bm{\sigma}_1-\bm{\sigma}_2}{2}\cdot\bm{P}\times\bm{k}$ & $C_1^*=\gamma_S-\gamma_T$ & $O^{*\prime\prime}_{{10}} = k^2\,\bm{P}\cdot\bm{\sigma}_1\,\bm{P}\cdot\bm{\sigma}_2$
& $D_{10}^*=-\frac{1}{4} (\gamma_S - \gamma_T) $ \\
$O^{*\prime}_{2} = \bm{\sigma}_1\cdot\bm{P}\bm{\sigma}_2\cdot\bm{Q}-\bm{\sigma_1}\cdot\bm{Q}\bm{\sigma}_2\bm{P}$ & $C_2^*=2 \gamma_T$ & $O^{*\prime\prime}_{{11}} = i Q^2(\bm{k}\times\bm{P}\cdot(\bm{\sigma}_1-\bm{\sigma}_2))$
& $ D_{11}^*=\frac{3}{2}( \gamma_S -\gamma_T)  -\frac{1}{2}(C_2- C_4 + C_5 - C_7)$ \\ 
$O^{*\prime}_{3} = P^2$ & $C_3^*=-\gamma_S$ & $O^{*\prime\prime}_{{12}} = i \bm{Q}\cdot\bm{P}(\bm{Q}\times\bm{k}\cdot(\bm{\sigma}_1-\bm{\sigma}_2))$
& $D_{12}^*=-\frac{1}{2} (C_5- C_7) $ \\
$O^{*\prime}_{4} = \bm{\sigma}_1\cdot\bm{\sigma}_2 P^2$ & $C_4^*=-\gamma_T$ & $O^{*\prime\prime}_{{13}} = i P^2(\bm{k}\times\bm{Q}\cdot(\bm{\sigma}_1+\bm{\sigma}_2))$
& $D_{13}^*=\frac{1}{4}(\gamma_S - 3 \gamma_T ) +\frac{1}{2} C_5 $ \\
$O^{*\prime}_{5} = \bm{\sigma}_1\cdot\bm{P}\bm{\sigma}_2\cdot\bm{P}$ & $C_5^*=0$ & $O^{*\prime\prime}_{{14}} = i  \bm{P}\cdot\bm{k}(\bm{Q}\times\bm{P}\cdot (\bm{\sigma}_1+\bm{\sigma}_2))$
& $D_{14}^*=\frac{1}{2} (\gamma_S - \gamma_T)  -\frac{1}{4}C_5 $ \\
\cline{1-2}
$O''_{1} = k^{4}$  & $D_1$ & $O^{*\prime\prime}_{{15}} =i  \bm{Q}\cdot\bm{P}(\bm{P}\times\bm{k} \cdot (\bm{\sigma}_1+\bm{\sigma}_2))$
& $D_{15}^*=-\gamma_T  - \frac{1}{4} C_5 $ \\
$O''_{2} = Q^{4}$ & $D_2$ & $O^{*\prime\prime}_{{16}} =i  P^2(\bm{k}\times\bm{P} \cdot (\bm{\sigma}_1-\bm{\sigma}_2))$
& $ D_{16}^*=\frac{5}{8}(  \gamma_S -  \gamma_T)  $ \\ 
$O''_{3} = k^{2} Q^{2}$ & $D_3$ & $O^{*\prime\prime}_{{17}} = Q^2(\bm{Q}\cdot\bm{\sigma}_1\,\bm{P}\cdot\bm{\sigma}_2-\bm{Q}\cdot\bm{\sigma}_2\,\bm{P}\cdot\bm{\sigma}_1)$
& $ D_{17}^*=6 \gamma_T  - (2  C_4 +  C_7)  $ \\
$O''_{4} = (\bm{k} \times \bm{Q})^{2}$ & $D_4$ & $O^{*\prime\prime}_{{18}} = P^2Q^2$
& $D_{18}^*=4 \gamma_S   -C_2  $ \\
$O''_{5} = k^{4}\bm{\sigma}_{1} \cdot \bm{\sigma}_{2}$ & $D_5$ & $O^{*\prime\prime}_{{19}} = (\bm{P}\cdot\bm{Q})^2$
& $D_{19}^*=4 \gamma_S   -C_2 $ \\
$O''_{6} = Q^{4}\bm{\sigma}_{1} \cdot \bm{\sigma}_{2}$ & $D_6$ &  $O^{*\prime\prime}_{{20}} = P^2Q^2\bm{\sigma}_1\cdot\bm{\sigma}_2$
& $D_{20}^*=4 \gamma_T  -C_4 $ \\
$O''_{7} = k^{2} Q^{2}\bm{\sigma}_{1} \cdot \bm{\sigma}_{2}$ & $D_7$ & $O^{*\prime\prime}_{{21}} = (\bm{P}\cdot\bm{Q})^2\bm{\sigma}_1\cdot\bm{\sigma}_2$
& $D_{21}^*=4 \gamma_T -C_4 $ \\
$O''_{8} = (\bm{k} \times \bm{Q})^{2}\bm{\sigma}_{1} \cdot \bm{\sigma}_{2}$ & $D_8$ & $O^{*\prime\prime}_{{22}} = Q^2\,\bm{P}\cdot\bm{\sigma}_1\,\bm{P}\cdot\bm{\sigma}_2$
& $D_{22}^*=-2 \gamma_T  $ \\
$O''_{9} = \frac{ik^2}{2}\left(\bm{\sigma}_{1}+\bm{\sigma}_{2}\right) \cdot(\bm{Q} \times \bm{k})$ & $D_9$ & $O^{*\prime\prime}_{{23}} = \bm{P}\cdot\bm{Q}(\bm{P}\cdot\bm{\sigma}_1\,\bm{Q}\cdot\bm{\sigma}_2+\bm{P}\cdot\bm{\sigma}_2\,\bm{Q}\cdot\bm{\sigma}_1)$
& $D_{23}^*=2 \gamma_T  -\frac{1}{2} C_7 $ \\
$O''_{{10}} = \frac{iQ^2}{2}\left(\bm{\sigma}_{1}+\bm{\sigma}_{2}\right) \cdot(\bm{Q} \times \bm{k})$ & $D_{10}$ & $O^{*\prime\prime}_{{24}} = P^2\,\bm{Q}\cdot\bm{\sigma}_1\,\bm{Q}\cdot\bm{\sigma}_2$
& $D_{24}^*=-2 \gamma_T -C_7 $ \\
$O''_{{11}} =  k^{2}\bm{\sigma}_{1} \cdot \bm{k}\bm{\sigma}_{2} \cdot \bm{k}$ & $D_{11}$ & $O^{*\prime\prime}_{{25}} = P^2(\bm{P}\cdot\bm{\sigma}_1\,\bm{Q}\cdot\bm{\sigma}_2-\bm{P}\cdot\bm{\sigma}_2\,\bm{Q}\cdot\bm{\sigma}_1)$
& $D_{25}^*=-\frac{5}{2} \gamma_T $ \\
$O''_{{12}} = Q^{2}\bm{\sigma}_{1} \cdot \bm{k}\bm{\sigma}_{2} \cdot \bm{k}$ & $ D_{12}$ & $O^{*\prime\prime}_{{26}} = P^4$
& $D_{26}^*=\gamma_S $ \\
$O''_{{13}} =  k^{2}\bm{\sigma}_{1} \cdot \bm{Q}\bm{\sigma}_{2} \cdot \bm{Q} $ & $D_{13}$ & $O^{*\prime\prime}_{{27}} = P^4\bm{\sigma}_1\cdot\bm{\sigma}_2$
& $D_{27}^*=\gamma_T  $ \\
$O''_{{14}} =  Q^{2}\bm{\sigma}_{1} \cdot \bm{Q}\bm{\sigma}_{2} \cdot \bm{Q}$ & $D_{14}$ & $O^{*\prime\prime}_{{28}} = P^2\,\bm{P}\cdot\bm{\sigma}_1\,\bm{P}\cdot\bm{\sigma}_2$
& $D_{28}^*=0$ \\
$O''_{{15}} = \bm{\sigma}_{1} \cdot(\bm{k} \times \bm{Q}) \,\bm{\sigma}_{2} \cdot(\bm{k} \times \bm{Q})$ &$D_{15}$ & &  \\
$O''_{{16}} = i\bm{k} \cdot \bm{Q}\, \bm{Q} \times \bm{P} \cdot\left(\bm{\sigma_{1}}-\bm{\sigma_{2}}\right)$ &$D_{16}$  & & \\
$O''_{{17}} = \bm{k} \cdot \bm{Q}\,(\bm{k} \times \bm{P}) \cdot\left(\bm{\sigma_{1}} \times \bm{\sigma_{2}}\right)$ &$D_{17}$  & & \\
   \bottomrule
  \end{tabular*}
\end{scriptsize}
\end{table*}

\section{Non-relativistic reduction}\label{sec:4}
The non-relativistic reduction of the $\tilde{O}_{i}$ up to terms of order $p^4$ can be obtained along the lines of Ref.~\cite{Girlanda:2010zz} (see also Ref.~\cite{Xiao:2018jot}). One  starts from the (positive-energy) relativistic field
\begin{equation}
\psi^{(+)}(x)=\int \frac{d^{3} \mathbf{p}}{(2 \pi)^{3}} \frac{m}{E_{\mathbf{p}}} b_{s}(\mathbf{p}) u^{(s)}(\mathbf{p}) \mathrm{e}^{-i p \cdot x},
\end{equation}

with normalizations,

\begin{eqnarray}
\left\{b_{s}(\mathbf{p}), b_{s^{\prime}}^{\dagger}(\mathbf{k})\right\}&=&\frac{E_{\mathbf{p}}}{m} \delta_{s s^{\prime}}(2 \pi)^{3} \delta^{(3)}(\mathbf{p}-\mathbf{k}), \\
\bar{u}^{(s)}(\mathbf{k}) u^{\left(s^{\prime}\right)}(\mathbf{k})&=&\delta_{s s^{\prime}},  
\end{eqnarray}

and expresses it in terms of the non-relativistic one,
\begin{equation}
N(x)=\int \frac{d^{3} \mathbf{p}}{(2 \pi)^{3}} \phi^{(s)} \tilde{b}_{s}(\mathbf{p}) \mathrm{e}^{-i p \cdot x},
\end{equation}

with $\phi^{(s)}$ a two-component spin doublet, and the operators $\tilde{b}_{s}(\mathbf{p}) \equiv \sqrt{m / E_{\mathbf{p}}} b_{s}(\mathbf{p})$, according to
\begin{equation}
\psi(x)=[\left(\begin{array}{l}
1 \\
0
\end{array}\right)-\frac{i}{2 m}\left(\begin{array}{c}
0 \\
\sigma \cdot \nabla
\end{array}\right)+\frac{1}{8 m^{2}}\left(\begin{array}{c}
\nabla^{2} \\
0
\end{array}\right)-\frac{3 i}{16 m^{3}}\left(\begin{array}{c}
0 \\
\sigma \cdot \nabla \nabla^{2}
\end{array}\right)+\frac{11}{128 m^{4}}\left(\begin{array}{c}
\nabla^{4} \\
0
\end{array}\right)] N(x)+O\left(q^{5}\right)\,.   
\end{equation}
Using these relations, the properties of Pauli matrices, partial integrations and the fields' equations of motion to eliminate time derivatives \cite{Grosse-Knetter:1993tae}, one can perform the non-relativistic expansion of the relativistic operators of Table \ref{tab:RelOp}. 
The resulting expressions are reported in Appendix~\ref{app}. The algebraic calculations were automatized with the software FORM \cite{https://doi.org/10.48550/arxiv.math-ph/0010025}.
By inspection of the non-relativistic expansions of our 54 relativistic operators, we find that only 26 linearly independent combinations survive\footnote{As a check of the algebraic manipulations, we recovered the expansions reported in Ref.~\cite{Xiao:2018jot}, except for some
typos involving the operators $O_{22}$, $\tilde O_{20}$ and $\tilde O_{31}$ of that reference (private communication by Yang Xiao). Away from the center-of-mass system, the expansions of the 40 relativistic operators of Ref.~\cite{Xiao:2018jot} contain only 25 independent combinations, suggesting that some operator is missing in the proposed list.}. Two of them start at order $O(p^0)$,
\begin{align}
    \tilde O_S &= O_S + \frac{1}{4m^2} \bigl( O^{*\prime}_1 -  O^{*\prime}_3 \bigr) + \frac{1}{16 m^4} \bigl( \frac{3}{8} O^{*\prime\prime}_1 + \frac{3}{4} O^{*\prime\prime}_4 + \frac{1}{4} O^{*\prime\prime}_5 + \frac{5}{4} O^{*\prime\prime}_6 -\frac{1}{4} O^{*\prime\prime}_7 -\frac{1}{4} O^{*\prime\prime}_8 + \frac{1}{4} O^{*\prime\prime}_9 \nonumber\\&-\frac{1}{4} O^{*\prime\prime}_{10} + \frac{3}{2} O^{*\prime\prime}_{11} + \frac{1}{4} O^{*\prime\prime}_{13} + \frac{1}{2} O^{*\prime\prime}_{14} + \frac{5}{8} O^{*\prime\prime}_{16} + 4 O^{*\prime\prime}_{18} +4 O^{*\prime\prime}_{19} + O^{*\prime\prime}_{26} \bigr),\\
    \tilde O_T &= O_T + \frac{1}{4 m^2} \bigl( -O^{*\prime}_1 +2O^{*\prime}_2 - O^{*\prime}_4 \bigr) + \frac{1}{16m^4} \bigl( -\frac{3}{8} O^{*\prime\prime}_1 +\frac{3}{2} O^{*\prime\prime}_3 + \frac{1}{4} O^{*\prime\prime}_4 + \frac{3}{4} O^{*\prime\prime}_5 -\frac{1}{4} O^{*\prime\prime}_6 +\frac{5}{4} O^{*\prime\prime}_7 \nonumber\\&+\frac{1}{4} O^{*\prime\prime}_8 - \frac{1}{4} O^{*\prime\prime}_9 +\frac{1}{4} O^{*\prime\prime}_{10} - \frac{3}{2} O^{*\prime\prime}_{11} - \frac{3}{4} O^{*\prime\prime}_{13} - \frac{1}{2} O^{*\prime\prime}_{14} -  O^{*\prime\prime}_{15} - \frac{5}{8} O^{*\prime\prime}_{16} + 6 O^{*\prime\prime}_{17} +4 O^{*\prime\prime}_{20} + 4 O^{*\prime\prime}_{21}\nonumber\\& - 2 O^{*\prime\prime}_{22} + 2 O^{*\prime\prime}_{23} - 2 O^{*\prime\prime}_{24} - \frac{5}{2} O^{*\prime\prime}_{25} +  O^{*\prime\prime}_{27} \bigr),
    \end{align}
    and contain the relativistic ${\bf P}$ dependent corrections to the LO operators $O_S$ and $O_T$, to which they reduce in the center-of-mass frame.
    Subsequently, there are 7 operators starting at $O(p^2)$,
    \begin{align}
        \tilde O'_1 &= O'_1 + \frac{1}{4m^2} \bigl( -\frac{1}{2} O^{*\prime\prime}_{1} -  O^{*\prime\prime}_{4} -   O^{*\prime\prime}_{6} \bigr),\\
        \tilde O'_2 &= O'_2 + \frac{1}{4m^2} \bigl( -\frac{1}{2} O^{*\prime\prime}_{11} - O^{*\prime\prime}_{18} -  O^{*\prime\prime}_{19}\bigr),\\
        \tilde O'_3 &= O'_3 + \frac{1}{4m^2} \bigl( \frac{1}{2} O^{*\prime\prime}_{1} -2 O^{*\prime\prime}_{3} -   O^{*\prime\prime}_{5} -  O^{*\prime\prime}_{7}\bigr),\\
        \tilde O'_4 &= O'_4 + \frac{1}{4m^2} \bigl( \frac{1}{2} O^{*\prime\prime}_{11} -2 O^{*\prime\prime}_{17} -  O^{*\prime\prime}_{20} -  O^{*\prime\prime}_{21}\bigr),\\
        \tilde O'_5 &= O'_5 + \frac{1}{4m^2} \bigl( \frac{1}{4} O^{*\prime\prime}_{2} +\frac{1}{4} O^{*\prime\prime}_{3} -\frac{1}{2} O^{*\prime\prime}_{11} - \frac{1}{2}  O^{*\prime\prime}_{12} + \frac{1}{2}  O^{*\prime\prime}_{13} - \frac{1}{4}  O^{*\prime\prime}_{14} - \frac{1}{4}  O^{*\prime\prime}_{15}\bigr),\\
        \tilde O'_6 &= O'_6 + \frac{1}{4m^2} \bigl(  O^{*\prime\prime}_{2} -  O^{*\prime\prime}_{8} - \frac{1}{2}  O^{*\prime\prime}_{9}\bigr),\\
        \tilde O'_7 &= O'_7 + \frac{1}{4 m^2} \bigl( \frac{1}{2} O^{*\prime\prime}_{11} +\frac{1}{2} O^{*\prime\prime}_{12} - O^{*\prime\prime}_{17} - \frac{1}{2} O^{*\prime\prime}_{23} - O^{*\prime\prime}_{24}\bigr),
    \end{align}
which reduce to the $O(p^2)$ operators $O'_{i=1,...,7}$ in the center-of-mass frame and contain in addition the relativistic drift corrections. 
  Each one of the above independent combinations will enter the Hamiltonian multiplied by a corresponding LEC. Since the tilded operators reduce to the untilded ones in the center-of-mass frame, the LECs must be the ones entering the standard NN contact potential up to $O(p^2)$ in Eqs.~(\ref{eq:h0}) and~(\ref{eq:h2}), respectively $C_S$, $C_T$ and $C_{i=1,...,7}$.
Finally there are 17 LECs multiplying the remaining $O(p^4)$ operators $O''_{i=1,...,17}$, including the ${\bf P}$-dependent operators $O_{16}$ and $O_{17}$, already identified in Ref.~\cite{Girlanda:2020pqn}. For example, in terms of the relativistic operators displayed in the Table \ref{tab:RelOp}, we have
\begin{align}
\frac{O''_{16}}{  m^4  } &=  - \tilde{O}_{10} - \frac{1}{2} \tilde{O}_{11} - \frac{3}{2} \tilde{O}_{12} + \tilde{O}_{13} -\tilde{O}_{16} + \frac{1}{2} \tilde{O}_{24} - \tilde{O}_{28} - 2 \tilde{O}_{32} +\tilde{O}_{33} -2\tilde{O}_{35} -2 \tilde{O}_{36} + 2 \tilde{O}_{37} + \frac{3}{2} \tilde{O}_{39} + \tilde{O}_{46},\\
 \frac{  O''_{17}}{ m^4 } &=  -4 \tilde{O}_{10} + 2 \tilde{O}_{11} -2  \tilde{O}_{12} +4 \tilde{O}_{13} - 2 \tilde{O}_{24} + 4 \tilde{O}_{28} -4 \tilde{O}_{33} +2 \tilde{O}_{39} .
    \end{align}
    Therefore the Hamiltonian density will take the following form, up to $O(p^4)$,
    \begin{equation}
        {\cal H}=C_S \tilde O_S + C_T \tilde O_T + \sum_{i=1}^7 C_i \tilde O'_i + \sum_{i=1}^{17} D_i O''_i.
        \end{equation}
Matching to  Eqs.(\ref{eq:h2}-\ref{eq:h4}) leads to the identifications of the LECs $C_i^*$ and $D_i^*$  reported in Table~\ref{tab:NRBase}, which extend the results of Refs.~\cite{Girlanda:2010ya, Pastore:2009is} for the $C_i^*$.

\section{Conclusions}\label{sec:5}
We have identified the constraints from relativity on the $O(p^4)$ NN contact interaction, starting from a manifestly Lorentz invariant effective Lagrangian. The results of this analysis were twofold. On one side, we have explicitly determined those LECs which parametrize the ${\bf P}$-dependent NN interaction representing "drift terms", i.e. the boosting of the interaction away from the center-of-mass frame.
On the other hand, we have confirmed the existence of two unconstrained LECs, whose effect vanishes in the NN center-of-mass frame and consequently cannot be determined from NN scattering data. Instead, these LECs play an important role in larger systems. As was shown in Ref.~\cite{Girlanda:2020pqn} these interactions (like others specifying off-shell effects)  are unitarily equivalent to specific combinations of the subleading three-nucleon contact operators of Ref.~\cite{Girlanda:2011fh} (see also Ref.~\cite{Reinert:2017usi}). Their effect on $p-d$ scattering observables has recently been investigated in Ref.~\cite{Girlanda:2023znc}, and found to be crucial for the resolution of the long-standing $A_y$ puzzle, since the corresponding LECs can be used as fitting parameters.
This contrasts with the LECs $C_i^*$ and $D_i^*$, whose magnitude we have determined in the present paper. Their impact in $A>2$ systems should be quantified accordingly. Work along these lines is left for future investigations.
 \appendix
\section{Non-relativistic expansions} \label{app}
 Here we list the explicit form of the non-relativistic expansions of the $\tilde{O}_i$ operators defined in Table \ref{tab:RelOp} in terms of  the basis operators of Table \ref{tab:NRBase}: 
{\allowdisplaybreaks
\begin{align}
  \tilde O_1 &= O_S + \frac{1}{ 4m^2} \bigl( -4 O'_2 +2 O'_5 
  + O^{*\prime}_1-  O^{*\prime}_{3}\bigr) +
  \frac{1}{16 m^4} \bigl( 16 O''_2 + 8 O''_3 -5 O''_4 - \frac{3}{2} O''_9 - 10 O''_{10} \nonumber \\
  &   - O''_{15} - \frac{1}{2} O''_{17} + \frac{3}{8} O^{*\prime\prime}_1 + \frac{1}{2} O^{*\prime\prime}_2  +  \frac{1}{32} O^{*\prime\prime}_3 + \frac{3}{4} O^{*\prime\prime}_4+ \frac{1}{4} O^{*\prime\prime}_5 + \frac{5}{4} O^{*\prime\prime}_6 - \frac{1}{4} O^{*\prime\prime}_7 - \frac{1}{4} O^{*\prime\prime}_8 \nonumber \\
  &   + \frac{1}{4} O^{*\prime\prime}_9 - \frac{1}{4}O^{*\prime\prime}_{10}  + \frac{5}{2} O^{*\prime\prime}_{11} -  O^{*\prime\prime}_{12} + \frac{5}{4} O^{*\prime\prime}_{13} - \frac{1}{2} O^{*\prime\prime}_{15} + \frac{5}{8} O^{*\prime\prime}_{16} + 8 O^{*\prime\prime}_{18} + 8 O^{*\prime\prime}_{19} +  O^{*\prime\prime}_{26} \bigr),\\
  \tilde O_2 &= \tilde O_1 + \frac{1}{4 m^2} \bigl(  O'_1 + 8 O'_2\bigr) +   \frac{1}{ 16 m^4} \bigl( -32O''_2 -8O''_3 +4 O''_4 + 2 O''_9 + 16O''_{10}  \nonumber \\
  &  - \frac{1}{2} O^{*\prime\prime}_{1} - O^{*\prime\prime}_{4} -  O^{*\prime\prime}_{6} - 4 O^{*\prime\prime}_{11} -8O^{*\prime\prime}_{18} - 8 O^{*\prime\prime}_{19} \bigr), \\
  \tilde O_3 &= -\frac{1}{4 m^2} O'_1  + \frac{1}{16 m^4} \bigl( 8 O''_3 - 4 O''_4 - 2 O''_9  + \frac{1}{2} O^{*\prime\prime}_{1} + O^{*\prime\prime}_{4} + O^{*\prime\prime}_{6} \bigr),\\
    \tilde O_4 &= \tilde O_1 + \frac{1}{4 m^2} \bigl( 2 O'_1 + 16 O'_2 \bigr) + \frac{1}{16m^4} \bigl(   O''_1
    + 8O''_4 + 4 O''_9 + 32 O''_{10} -  O^{*\prime\prime}_{1} - 2 O^{*\prime\prime}_{4} - 2O^{*\prime\prime}_{6} - 8 O^{*\prime\prime}_{11}\nonumber\\& - 16 O^{*\prime\prime}_{18} - 16 O^{*\prime\prime}_{19} \bigr) , \\
    \tilde O_5 &= \tilde O_3 + \frac{1}{16 m^4} \bigl( -  O''_1 - 8 O''_3 \bigr),\\
    \tilde O_6 &= \frac{1}{16 m^4} O''_1 ,\\
    \tilde O_7 &= \tilde O_1 + \frac{1}{4m^2} \bigl( 8 O'_2 -4 O'_5 \bigr) + \frac{1}{16m^4} \bigl( -32 O''_2 - 8 O''_3 + 4 O''_4 + 2 O''_9 +24 O''_{10} +4 O''_{15} + 2 O''_{16} + O''_{17} \nonumber\\& -  O^{*\prime\prime}_{2}  -  O^{*\prime\prime}_{3}- 2 O^{*\prime\prime}_{11} + 2 O^{*\prime\prime}_{12} - 2 O^{*\prime\prime}_{13} +  O^{*\prime\prime}_{14} +  O^{*\prime\prime}_{15} - 8O^{*\prime\prime}_{18} -8O^{*\prime\prime}_{19} \bigr) ,\\
\tilde O_8 &= \tilde O_7  + \frac{1}{4 m^2} \bigl(  O'_1 + 8 O'_2\bigr) + \frac{1}{16 m^4} \bigl( 32 O''_2    + 4 O''_4 - 2 O''_9 - 16 O''_{10} -\frac{1}{2} O^{*\prime\prime}_{1} - O^{*\prime\prime}_{4}
- O^{*\prime\prime}_{6} - 4 O^{*\prime\prime}_{11}\nonumber\\& - 8 O^{*\prime\prime}_{18} -8 O^{*\prime\prime}_{19}\bigr),\\
\tilde O_9 &= -\frac{1}{4 m^2} O'_1 + \frac{1}{16 m^4} \bigl( -4O''_4 + 2 O''_9 + \frac{1}{2} O^{*\prime\prime}_{1} +O^{*\prime\prime}_{4} +  O^{*\prime\prime}_{6} \bigr) ,\\
\tilde O_{10} &= \frac{1}{m^2} O'_5 + \frac{1}{16 m^4} \bigl( -4  O''_4 + 2 O''_9 - 8 O''_{10} - 4 O''_{15} - 2 O''_{16} -  O''_{17} +  O^{*\prime\prime}_{2} +  O^{*\prime\prime}_{3} - 2 O^{*\prime\prime}_{11} - 2 O^{*\prime\prime}_{12 } \nonumber\\&+ 2 O^{*\prime\prime}_{13} -  O^{*\prime\prime}_{14} -O^{*\prime\prime}_{15} \bigr),\\
\tilde O_{11} &= \tilde O_{10} + \frac{1}{16m^4} \bigl( 4 O''_9 + 32 O''_{10} \bigr) ,\\
\tilde O_{12} &= -\frac{1}{4 m^4} O''_9,\\
\tilde O_{13} &= \frac{1}{4 m^2} O'_6 + \frac{1}{16m^4} \bigl( 2 O''_8 - O''_{11} - 6 O''_{12} -2O''_{13} - 2 O''_{15} + O''_{17} +
  O^{*\prime\prime}_{2} -  O^{*\prime\prime}_{8} -  \frac{1}{2} O^{*\prime\prime}_{9} \bigr) ,\\
\tilde O_{14} &= \tilde O_{13} + \frac{1}{16m^4} \bigl(  O''_{11} + 8 O''_{12} \bigr), \\
\tilde O_{15} &= -\frac{1}{16 m^4} O''_{11},\\
\tilde O_{16} &= \frac{1}{ 16 m^4} \bigl( -8 O''_8 -8 O''_{12} + 8 O''_{13} + 8 O''_{15} \bigr),\\
\tilde O_{17} &= \tilde O_1 + \frac{1}{4 m^2} \bigl( -O'_1 + 8 O'_2 - O'_3 - 8 O'_5 + O'_6 \bigr) \nonumber \\
& +\frac{1}{16m^4}\bigl( \ O''_{1} - 32 O''_{2} -8 O''_{3} + 4 O''_{4} +  O''_{5} + 8 O''_{7} -2O''_{8} + 8 O''_{9} + 32O''_{10} - O''_{11} - 6 O''_{12} - 2 O''_{13}\nonumber\\& - 2 O''_{15} + 4 O''_{16} - O''_{17} -  O^{*\prime\prime}_{2} + O^{*\prime\prime}_{4} + O^{*\prime\prime}_{5} + O^{*\prime\prime}_{6} +  O^{*\prime\prime}_{7} - O^{*\prime\prime}_{8} -  \frac{1}{2} O^{*\prime\prime}_{9} + 4O^{*\prime\prime}_{12}\nonumber\\& - 4O^{*\prime\prime}_{13} +2 O^{*\prime\prime}_{14} + 2O^{*\prime\prime}_{15} - 8O^{*\prime\prime}_{18} - 8O^{*\prime\prime}_{19} \bigr),
\\
\tilde O_{18} &=\tilde O_{17} + \frac{1}{4m^2} \bigl( O'_1 + 8 O'_2 \bigr) + \frac{1}{16m^4} \bigl( -O''_1 + 32 O''_2 -8 O''_3 + 4 O''_4 -  O''_5 - 8 O''_7 -6 O''_9 - 48 O''_{10} +  O''_{11} \nonumber\\&+8 O''_{12} -\frac{1}{2} O^{*\prime\prime}_{1} - O^{*\prime\prime}_{4} - O^{*\prime\prime}_{6} -4 O^{*\prime\prime}_{11} - 8O^{*\prime\prime}_{18} - 8 O^{*\prime\prime}_{19} \bigr) , \\
\tilde O_{19} &= -\frac{1}{4 m^2} O'_1 + \frac{1}{16 m^4} \bigl(  O''_1 - 4 O''_4 +  O''_5 + 6O''_9 -  O''_{11} + \frac{1}{2} O^{*\prime\prime}_{1} +  O^{*\prime\prime}_{4} +  O^{*\prime\prime}_{6} \bigr) , \\
\tilde O_{20} &= \tilde O_1 + \frac{1}{4m^2} \bigl( 16 O'_2 - 8 O'_5 \bigr) + \frac{1}{16m^4} \bigl( -16 O''_3 + 8 O''_4 + 4O''_9 - 16 O''_{10} - 8 O''_{15} + 4 O''_{16} + 2 O''_{17} -2 O^{*\prime\prime}_{2} \nonumber\\& - 2 O^{*\prime\prime}_{3} - 4 O^{*\prime\prime}_{11} + 4 O^{*\prime\prime}_{12}- 4 O^{*\prime\prime}_{13} + 2 O^{*\prime\prime}_{14} +2 O^{*\prime\prime}_{15} - 16O^{*\prime\prime}_{18} - 16O^{*\prime\prime}_{19} \bigr), \\
\tilde O_{21} &= \tilde O_{20} + \frac{1}{4m^2} \bigl( O'_1 + 8 O'_2 \bigr) + \frac{1}{16m^4} \bigl(96 O''_2 + 8 O''_3 + 4 O''_4 - 6 O''_9 - 48 O''_{10} - \frac{1}{2} O^{*\prime\prime}_{1} - O^{*\prime\prime}_{4} -  O^{*\prime\prime}_{6} \nonumber\\&- 4 O^{*\prime\prime}_{11} - 8O^{*\prime\prime}_{18} - 8O^{*\prime\prime}_{19}\bigr), \\
\tilde O_{22} &= - \frac{1}{4 m^2} O'_1 + \frac{1}{16m^4} \bigl( - 8 O''_3 - 4 O''_4 + 6 O''_9 + \frac{1}{2} O^{*\prime\prime}_{1} +  O^{*\prime\prime}_{4} + O^{*\prime\prime}_{6} \bigr) , \\
\tilde O_{23} &= \frac{1}{16 m^4} \bigl( 8O''_8 - 8 O''_{12} + 8 O''_{13} - 8 O''_{15} \bigr), \\
\tilde O_{24} &= \tilde O_{11} +\frac{1}{16 m^4} \bigl( -4O''_9 +16 O''_{15} \bigr), \\
\tilde O_{25} &= -\tilde O_{13} + \frac{1}{4 m^2} O'_3 + \frac{1}{16m^4} \bigl( 4 O''_8 +2O''_9 -  O''_{11} - 8 O''_{12} - 8 O''_{15} + 4 O''_{17} + \frac{1}{2} O^{*\prime\prime}_{1} - 2 O^{*\prime\prime}_{3} - \ O^{*\prime\prime}_{5} -  O^{*\prime\prime}_{7} \bigr), \\
\tilde O_{26} &= \tilde O_{25} + \frac{1}{16m^4} \bigl( O''_5 + 8 O''_7 -  O''_{11} - 8O''_{12} \bigr), \\
\tilde O_{27} &= \frac{1}{16m^4} \bigl( - O''_5 + O''_{11} \bigr), \\
\tilde O_{28} &= \tilde O_{25} + \tilde O_{10} + \frac{1}{16m^4} \bigl( -8 O''_7 - 4 O''_9 + 8O''_{12} + 16O''_{15} \bigr), \\
\tilde O_{29} &= \tilde O_{28} + \frac{1}{16m^4} \bigl(  O''_5 + 8 O''_7 +4 O''_9 + 32 O''_{10} -  O''_{11} - 8O''_{12} \bigr), \\
\tilde O_{30} &= \frac{1}{16 m^4} \bigl( - O''_5 -4 O''_9 + O''_{11} \bigr), \\
\tilde O_{31} &= - O_T + \frac{1}{4m^2} \bigl(4 O'_4 - 2 O'_5 + O'_6 - 8 O'_7 +O^{*\prime}_1 -2O^{*\prime}_2 + O^{*\prime}_4\bigr) \nonumber \\
& + \frac{1}{16m^4} \bigl(  O''_4 - 16O''_6 -8O''_7 + 2 O''_8 + \frac{3}{2} O''_9 + 10 O''_{10} -  O''_{11} - 2 O''_{12}+ 10 O''_{13} + 32 O''_{14} + 3 O''_{15} + 2O''_{16}\nonumber\\& -\frac{3}{2} O''_{17} + \frac{3}{8} O^{*\prime\prime}_{1} + \frac{1}{2} O^{*\prime\prime}_{2} -2 O^{*\prime\prime}_{3} - \frac{1}{4} O^{*\prime\prime}_{4} - \frac{3}{4} O^{*\prime\prime}_{5} + \frac{1}{4} O^{*\prime\prime}_{6}- \frac{5}{4} O^{*\prime\prime}_{7} - \frac{5}{4} O^{*\prime\prime}_{8} -\frac{1}{4} O^{*\prime\prime}_{9} -\frac{1}{4} O^{*\prime\prime}_{10} \nonumber\\&+ \frac{1}{2} O^{*\prime\prime}_{11} -3 O^{*\prime\prime}_{12} - \frac{1}{4} O^{*\prime\prime}_{13} +  O^{*\prime\prime}_{14} + \frac{3}{2} O^{*\prime\prime}_{15} + \frac{5}{8} O^{*\prime\prime}_{16}-6 O^{*\prime\prime}_{17}-8 O^{*\prime\prime}_{20} -8 O^{*\prime\prime}_{21}+2 O^{*\prime\prime}_{22} + 2 O^{*\prime\prime}_{23} + 10 O^{*\prime\prime}_{24} \nonumber\\&+ \frac{5}{2} O^{*\prime\prime}_{25} -  O^{*\prime\prime}_{27} \bigr), \\
\tilde O_{32} &= \tilde O_{31} + \frac{1}{4m^2} \bigl( - O'_3 - 8 O'_4 \bigr) + \frac{1}{16m^4} \bigl( 32O''_6 + 8O''_7 - 4 O''_8 -2 O''_9 -16 O''_{10} + O''_{11} + 8 O''_{12} - 8 O''_{13} - 64 O''_{14} \nonumber\\&-\frac{1}{2} O^{*\prime\prime}_{1} + 2 O^{*\prime\prime}_{3} + O^{*\prime\prime}_{5} +O^{*\prime\prime}_{7}
-4O^{*\prime\prime}_{11} + 16 O^{*\prime\prime}_{17} + 8 O^{*\prime\prime}_{20} + 8 O^{*\prime\prime}_{21} \bigr), \\
\tilde O_{33} &= \frac{1}{4 m^2} O'_3 + \frac{1}{16m^4} \bigl( -8O''_7 +4 O''_8 + 2 O''_9 - O''_{11} + 8 O''_{13} + \frac{1}{2} O^{*\prime\prime}_{1} - 2 O^{*\prime\prime}_{3} - O^{*\prime\prime}_{5} - O^{*\prime\prime}_{7} \bigr), \\
\tilde O_{34} &= \tilde O_{31} + \frac{1}{4m^2} \bigl( -2O'_3 - 16 O'_4\bigr) + \frac{1}{16m^4} \bigl( - O''_5 - 8O''_8 -4O''_9 - 32O''_{10} + 2 O''_{11} + 16 O''_{12} -16  O''_{13} \nonumber\\& - 128 O''_{14} -O^{*\prime\prime}_{1}+ 4 O^{*\prime\prime}_{3} +2O^{*\prime\prime}_{5} + 2O^{*\prime\prime}_{7}-8O^{*\prime\prime}_{11} + 32 O^{*\prime\prime}_{17} + 16 O^{*\prime\prime}_{20} +16  O^{*\prime\prime}_{21} \bigr), \\
\tilde O_{35} &= \tilde O_{33} + \frac{1}{16m^4} \bigl(  O''_5 + 8 O''_7 \bigr), \\
\tilde O_{36} &= -\frac{1}{16 m^4} O''_5, \\
\tilde O_{37} &= -\frac{4}{ m^2} O'_7 + \frac{1}{16 m^4} \bigl( -8 O''_8 + 8 O''_{12} + 8 O''_{13} + 8O''_{15} +8 O''_{16} - 8 O^{*\prime\prime}_{11} -8 O^{*\prime\prime}_{12} + 16 O^{*\prime\prime}_{17} + 8 O^{*\prime\prime}_{23} + 16 O^{*\prime\prime}_{24} \bigr), \\
\tilde O_{38} &= \tilde O_{37} + \frac{1}{16m^4} \bigl( -16O''_{13} - 128 O''_{14} \bigr), \\
\tilde O_{39} &= \frac{1}{m^4} O''_{13}, \\
\tilde O_{40} &= -\tilde O_{37} +\tilde O_{10} -\frac{4}{m^2} O'_4 + \frac{1}{16 m^4} \bigl( -8O''_8 + 8 O''_{12} + 8 O''_{13} + 8O''_{15} + 8O''_{16} -8O^{*\prime\prime}_{11} + 32O^{*\prime\prime}_{17} + 16O^{*\prime\prime}_{20} + 16O^{*\prime\prime}_{21}\bigr), \\
\tilde O_{41} &= \tilde O_{40} + \frac{1}{16m^4} \bigl( -128 O''_6 - 16O''_7 +4 O''_9 + 32 O''_{10} + 16O''_{13} + 128 O''_{14} \bigr), \\
\tilde O_{42} &= \frac{1}{16 m^4} \bigl(16 O''_7 - 4 O''_9 - 16O''_{13} \bigr), \\
\tilde O_{43} &= 2 \tilde O_{31} -\frac{1}{2 m^2} O'_6 + \frac{1}{16m^4} \bigl( 4 O''_8 + 2 O''_{11} + 4O''_{12} - 4O''_{13} - 4O''_{15} + 2 O''_{17} - 2 O^{*\prime\prime}_{2} + 2 O^{*\prime\prime}_{8} +  O^{*\prime\prime}_{9} \bigr), \\
\tilde O_{44} &=  \tilde O_{43} +\frac{1}{4m^2} \bigl( -2 O'_3 - 16O'_4 \bigr)  + \frac{1}{16m^4} \bigl( 64O''_6 + 16O''_7 -8 O''_8 -4 O''_9 -32 O''_{10} -16 O''_{13} - 128 O''_{14}\nonumber\\&  - O^{*\prime\prime}_{1} + 4 O^{*\prime\prime}_{3} + 2 O^{*\prime\prime}_{5}+ 2 O^{*\prime\prime}_{7}-8 O^{*\prime\prime}_{11} + 32 O^{*\prime\prime}_{17} + 16O^{*\prime\prime}_{20} +16 O^{*\prime\prime}_{21} \bigr), \\
\tilde O_{45} &= \frac{1}{2 m^2} O'_3 + \frac{1}{16m^4} \bigl( -16 O''_7 + 8 O''_8 + 4 O''_9 + 16O''_{13} +  O^{*\prime\prime}_{1} - 4 O^{*\prime\prime}_{3} -2 O^{*\prime\prime}_{5} -2O^{*\prime\prime}_{7} \bigr), \\
\tilde O_{46} &= \tilde O_{43} -2 \tilde O_{37} + 2 \tilde O_{10} -\frac{4}{m^2} O'_4 + \frac{1}{16m^4} \bigl( 64 O''_6 + 16O''_7 -16 O''_8 -8 O''_9 -32 O''_{10} + 16O''_{12} - 16O''_{13} \nonumber\\&- 128 O''_{14} +16 O''_{15} +16 O''_{16} - 8O^{*\prime\prime}_{11} + 32 O^{*\prime\prime}_{17}+16 O^{*\prime\prime}_{20}+ 16O^{*\prime\prime}_{21} \bigr), \\
\tilde O_{47} &= \tilde O_{46}  +\frac{1}{4m^2} \bigl( -2 O'_3 - 16 O'_4 \bigr) + \frac{1}{16m^4} \bigl( - 64 O''_6 -8 O''_8 + 4 O''_9 + 32 O''_{10} + 16O''_{13} +128 O''_{14} - O^{*\prime\prime}_{1} + 4 O^{*\prime\prime}_{3}\nonumber\\&  + 2 O^{*\prime\prime}_{5}+2O^{*\prime\prime}_{7} -8 O^{*\prime\prime}_{11} + 32 O^{*\prime\prime}_{17} +16 O^{*\prime\prime}_{20} + 16O^{*\prime\prime}_{21} \bigr), \\
\tilde O_{48} &= \tilde O_{45} +\frac{1}{16m^4} \bigl(16 O''_7 - 8 O''_9 - 32 O''_{13} \bigr),\\
\tilde O_{49} &= -2 \tilde O_{31} + \frac{1}{4m^2} \bigl( -2O'_1 - 2 O'_3 + 16O'_4 -16 O'_5 +4 O'_6 - 32 O'_7 \bigr) + \frac{1}{16m^4} \bigl( 2O''_1 +16 O''_3 -8O''_4 + 2 O''_5 \nonumber\\&- 64O''_6 -16 O''_7 + 16O''_9 + 64 O''_{10} -4O''_{11} - 8O''_{12} +40 O''_{13} + 128 O''_{14} + 8 O''_{15} -8O''_{16}  + 2 O^{*\prime\prime}_{4} + 2 O^{*\prime\prime}_{5}  \nonumber\\&+ 2O^{*\prime\prime}_{6}+ 2O^{*\prime\prime}_{7} - 4O^{*\prime\prime}_{8} - 2 O^{*\prime\prime}_{9} -8 O^{*\prime\prime}_{12} -8 O^{*\prime\prime}_{13} +4 O^{*\prime\prime}_{14} + 4O^{*\prime\prime}_{15} -16 O^{*\prime\prime}_{20} - 16O^{*\prime\prime}_{21}+ 16 O^{*\prime\prime}_{23} + 32 O^{*\prime\prime}_{24} \bigr), \\
\tilde O_{50} &=  \tilde O_{49} +\frac{1}{4m^2} \bigl( 2 O'_3 + 16 O'_4 \bigr) + \frac{1}{16m^4} \bigl( -2 O''_1 -16 O''_3 - 2 O''_5 + 64 O''_6 -16 O''_7 + 8O''_8 - 12 O''_9 - 96 O''_{10} \nonumber\\& + 2O''_{11} + 16O''_{12} - 16O''_{13}- 128 O''_{14} + O^{*\prime\prime}_{1} - 4 O^{*\prime\prime}_{3} -2O^{*\prime\prime}_{5} - 2O^{*\prime\prime}_{7} +8 O^{*\prime\prime}_{11} - 32 O^{*\prime\prime}_{17} - 16O^{*\prime\prime}_{20} - 16O^{*\prime\prime}_{21} \bigr), \\
\tilde O_{51} &= -\frac{1}{2 m^2} O'_3 + \frac{1}{16m^4} \bigl( 2 O''_1 +2 O''_5 - 8 O''_8 + 12 O''_9 -2 O''_{11} +16 O''_{13} -O^{*\prime\prime}_{1} +4 O^{*\prime\prime}_{3} + 2O^{*\prime\prime}_{5} +2 O^{*\prime\prime}_{7} \bigr), \\
\tilde O_{52} &= \frac{1}{4m^2} \bigl( -O'_1 + 16O'_4 - 8 O'_5 - 16O'_7 \bigr) + \frac{1}{16m^4} \bigl( -4 O''_4 -16 O''_7 + 8 O''_8 + 2 O''_9 -16 O''_{10} + 8O''_{12} \nonumber\\&+ 8O''_{13} -4 O''_{16} + 2 O''_{17} + \frac{1}{2} O^{*\prime\prime}_{1} -2 O^{*\prime\prime}_{2} -2 O^{*\prime\prime}_{3} + O^{*\prime\prime}_{4} +O^{*\prime\prime}_{6} +4 O^{*\prime\prime}_{11} -4O^{*\prime\prime}_{12} -4 O^{*\prime\prime}_{13} + 2O^{*\prime\prime}_{14} + 2O^{*\prime\prime}_{15} \nonumber\\&- 16 O^{*\prime\prime}_{17}- 16O^{*\prime\prime}_{20} - 16O^{*\prime\prime}_{21} + 8 O^{*\prime\prime}_{23} + 16O^{*\prime\prime}_{24} \bigr), \\
\tilde O_{53} &= \tilde O_{52} + \frac{1}{16m^4} \bigl( - O''_1 - 8O''_3 + 128 O''_6 + 16O''_7 - 8 O''_9 - 64 O''_{10} - 16O''_{13} - 128 O''_{14} \bigr), \\
\tilde O_{54} &=\frac{1}{16 m^4} \bigl(O''_1 -16 O''_7 + 8O''_9 +16 O''_{13} \bigr).
\end{align}
}

\bibliographystyle{elsarticle-num}

\bibliography{p4refs}
\end{document}